\documentclass[aps,amsmath,amssymb,twocolumn,nofootinbib]{revtex4}
\usepackage{graphicx}
\usepackage{pgfplots}
\pgfplotsset{compat=1.17}
\usepackage{dcolumn}
\usepackage{color}
\usepackage{scrextend}
\usepackage{subfig}
\usepackage{bm}
\usepackage{amsmath,amssymb,graphicx}
\usepackage{epsfig}
\usepackage{enumerate}
\usepackage{array}
\usepackage{multirow}
\usepackage{ragged2e}
\usepackage{textgreek}
\usepackage[normalem]{ulem}
\usepackage{xcolor}
\usepackage{tikz}
\usetikzlibrary{arrows.meta,positioning,fit,calc}
\usepackage{subcaption}
 \usetikzlibrary{arrows.meta,positioning,fit,calc}
\usepackage{lipsum}
\usepackage{ntheorem}
\usepackage{xcolor}
\usepackage{hyperref}

\usepackage{pgfplots}
\usepgfplotslibrary{groupplots}
\pgfplotsset{compat=1.18}

\definecolor{deepteal}{RGB}{0,102,102}
\definecolor{burntorange}{RGB}{184,102,37}
\definecolor{slateblue}{RGB}{90,98,140}
\definecolor{deepmaroon}{RGB}{128,48,64}

\pgfplotsset{
  myaxis/.style={
    width=0.305\textwidth,
    height=0.24\textwidth,
    thick,
    tick style={black},
    label style={font=\small},
    ticklabel style={font=\small},
    title style={font=\small},
    xminorticks=false,
    yminorticks=false
  }
}

\begin{document}
\title{Exceptional Points as Witnesses of Non-Gaussian Noise}
\author{Aritra Ghosh$^1$\footnote{aritraghosh500@gmail.com}, Subhashish Banerjee$^2$, and M. Bhattacharya$^1$}
\affiliation{$^1$School of Physics and Astronomy, Rochester Institute of Technology, 84 Lomb Memorial Drive, Rochester, New York 14623, USA\\
$^2$Department of Physics, Indian Institute of Technology Jodhpur, Rajasthan 342030, India}
\vskip-2.8cm
\date{\today}
\vskip-0.9cm

\begin{abstract}
Efficient control of quantum devices requires the knowledge of their environmental noise, yet the commonly measured power spectrum does not fully characterize non-Gaussian fluctuations of realistic environments. Here we show that exceptional points (EPs) in the noise-averaged dephasing response can expose observable signatures of non-Gaussian statistics, distinguishing environments with identical lower-order statistics. We also introduce a theoretical model that describes dephasing environments in terms of identical independent noise sources while preserving the power spectrum. Given a calibrated single-source model, the largest EP order extracted from the measured dephasing response determines the number of sources and predicts the higher-order statistics of the combined noise. We illustrate an atomtronic readout that maps distinct EP signatures onto an evolving ensemble-averaged atomic interference pattern. Our results connect non-Hermitian dynamics to environmental noise statistics, providing a route to characterize noise in quantum devices used in computing, metrology, and precision measurement.
\end{abstract}
\maketitle

\begin{figure}[t]
\centering
\includegraphics[width=0.88\columnwidth]{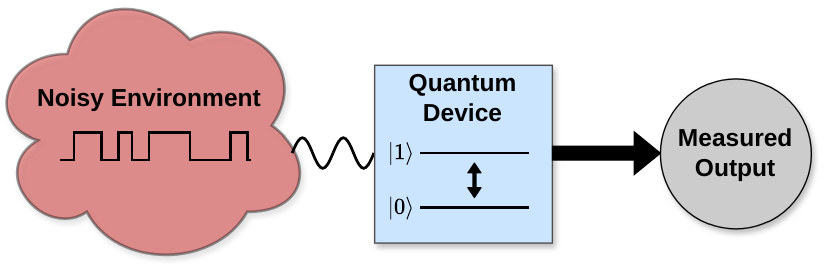}
\caption{\justifying{
Schematic of environmental noise coupled to a quantum system whose output is monitored. The fluctuations can degrade device performance while also imprinting statistical information on the measured response}, including signatures of non-Gaussian statistics.}
\label{fig1}
\end{figure}

\textit{Introduction.} Environmental fluctuations limit quantum coherence, but their statistics can also reveal the microscopic processes responsible for decoherence \cite{Ithier_2005,Paladino_2014,Galperin_2006}. As schematized in Fig. \ref{fig1}, controlled quantum systems can probe noise spectra \cite{Cywinski_2008,Bylander_2011,Yan_2012,Alvarez_2011,Yuge_2011,Young_2012,PazSilva_2014} and experiments have linked individual defects, charge-parity switching, and quasiparticles to variations in device performance \cite{Riste_2013,Gustavsson_2016,Klimov_2018,Serniak_2018,Schlor_2019,Pan_2022}. Noisy effects are relevant across diverse platforms, including superconducting qubits \cite{Bylander_2011,Riste_2013,Klimov_2018,Schlor_2019,Pan_2022,Sung_2019}, trapped ions \cite{Kotler_2013}, semiconductor quantum-dot devices \cite{Ye_2025}, solid-state spin memories \cite{Chen-noise_2018}, optical lattice clocks \cite{Al-Masoudi_2015}, quantum processors \cite{Maciejewski_2020,Ollitrault_2020}, and gravitational-wave parameter estimation \cite{Legin_2025}.

While for stationary classical Gaussian noise, the mean and two-point correlation determine all its statistics, non-Gaussian processes can share these quantities and yet produce different quantum dynamics because their higher correlations differ \cite{Benedetti_2013}. Non-Gaussian noise spectroscopy describes higher-order correlations
and spectra \cite{Sung_2019,Kotler_2013,Norris_2016,Ramon_2019,Wang_2020,Steven_2026,Curtis_2026,Message_2026}. Complementary studies address non-Gaussian noise and non-Markovian dynamics in open quantum systems \cite{Funo_2024,Tiwari_2026,Lin_2026,Kumar_2018,Kuffer_2025}. Random finite-state switching, realized via a so-called fluctuator, provides a useful effective description of non-Gaussian environments \cite{Schriefl_2006,Bergli_2009,Kuopanportti_2008,Wold_2012}, including dephasing \cite{Ramon_2015}, without requiring that a classical fluctuator represent a complete equilibrium reservoir \cite{Aharony_2019,Dattagupta_2022}.

On the other hand, exceptional points (EPs) have emerged as a rich and exciting feature of open quantum systems \cite{Minganti_2019,Arkhipov_2020,Zhong_2020,Sun_2024,Lin_2025}. At an EP, multiple eigenvalues and eigenvectors coalesce and the non-Hermitian dynamical generator becomes non-diagonalizable \cite{Heiss_2012,Miri_2019,Ozdemir_2019,Ashida_2020}. The EP order $r$ is defined to be the size of the largest Jordan block of the non-diagonalizable generator at the coalesced eigenvalue \cite{Prosen_2010,Wiersig_2025}. Such a block can produce an exponential multiplied by a polynomial in time of degree at most $r-1$ \cite{Ashida_2020}. Recently, EPs have also been studied in stochastic processes \cite{Li_2023,Pal_2026,Huang_2026}, and a magnetometer experiment has related a telegraph-noise response to a Liouvillian EP \cite{Geng_2026}. Stochastic Liouville theory connects a quantum probe's response to the
underlying noisy switching dynamics, not uniquely specified by the power spectrum \cite{Anderson_1954,Kubo_1963}. If EP signatures exist in this response, producing characteristic time transients \cite{Ashida_2020}, they may be strong candidates for probing environmental noise statistics. 

This work provides, to our knowledge, the first exact demonstration that EPs identified from the noise-averaged dephasing response of a quantum device can distinguish, through their order and location in parameter space, distinct stationary noise environments with the same mean, power spectrum, and all third-order correlations. Matching the third order provides a stronger test by ruling out third-order differences. For certain symmetric environments, the third-order correlations vanish identically \cite{Ramon_2019}, making fourth-order statistics the first general distinction beyond the mean and power spectrum. Motivated by fluctuator models of qubit dephasing
\cite{Galperin_2006,Schriefl_2006,Bergli_2009} and charge-parity switching \cite{Riste_2013}, we represent environmental noise by stationary finite-state Markov processes. For identical independent sources at fixed power spectrum, a known single-source model lets us infer higher-order noise statistics from the experimentally observable EP order and provides a theoretical benchmark for fitting and interpreting coherence measurements in platforms such as superconducting
\cite{Bylander_2011,Sung_2019} and nuclear-spin qubits \cite{Chen-noise_2018}. We finally use an atomtronic-ring model to illustrate how the predicted signal could be read from an atomic interference pattern.

\textit{Exceptional points and noise.}
Imagine changing a quantum system's dephasing environment while preserving both the mean and the measured power spectrum. For stationary classical Gaussian noise, the coherence decay would then remain unchanged. Discrete switching need not behave this way, since the spectrum records second-order temporal memory, yet it can miss how successive switching events are organized \cite{Ramon_2019}. We use the quantum system as a sensor by preparing it in a superposition state and monitoring its coherence over time. In our proposal we use units with $\hbar=1$ throughout and assume that the stationary noise is statistically independent of the quantum system when it is prepared. We consider pure dephasing with strength $g$, i.e., $H(t)=H_0+g\eta(t)V$, with $[H_0,V]=0$. Here we take $H_0$ and $V$ to be Hermitian, so each noise realization is unitary, thereby allowing us to focus only on non-Hermiticity due to stochasticity.

The noise is generated by a finite-state Markov system, i.e., its internal state changes according to a transition generator $W$ and each state has an associated measured noise value. Let $\boldsymbol\pi$ be the stationary probability column vector, $X$ be the diagonal matrix of noise values, and $\mathbf{1}^T=(1,1,\ldots,1)$. In the common eigenbasis $\{|m\rangle\}$ of $H_0$ and $V$, let $H_0|m\rangle=E_m|m\rangle$ and $V|m\rangle=v_m|m\rangle$. For a noise realization $\eta$, the density matrix evolves as $\rho_\eta(t)=U_\eta(t)\rho(0)U_\eta^\dagger(t)$, where $U_\eta(t)=\exp[-i\int_0^t H(t') dt']$. The coherence matrix element $\rho_{mn}^{(\eta)}(t)=\langle m|\rho_\eta(t)|n\rangle$
therefore satisfies $\rho_{mn}^{(\eta)}(t)=\rho_{mn}(0)e^{-i(E_m-E_n)t-i\chi\int_0^t\eta(t') dt'}$, with the coupling $\chi=g(v_m-v_n)$. After removing the deterministic phase coming from $H_0$ and averaging over the noise, this coherence is multiplied by the factor \cite{Kubo_1963,Fitzsimmons_1999}
\begin{equation}
\Phi(\chi,t)
=\mathbf{1}^T e^{(W-i\chi X)t}\boldsymbol\pi .
\label{coherence-factor}
\end{equation}
We denote by $G(\chi)=W-i\chi X$ the tilted generator. Each entry of $e^{G(\chi)t}\boldsymbol\pi$ gives the average phase factor for trajectories ending in one noise state, multiplied by the probability of that state. Summing these entries gives $\Phi$. For example, three-state noise gives three coupled phase averages, governed by a $3\times3$ matrix $G$, which can have an EP$_3$ even when the quantum system is a qubit with a two-dimensional Hilbert space. We call an order-$r$ EP at eigenvalue $\lambda$ ``coherence-visible'' when $\Phi$ contains a nonzero $t^{r-1}e^{\lambda t}$ term. Equivalently, its Laplace transform $\widehat{\Phi}(\chi,s)=\mathbf{1}^T[sI-G(\chi)]^{-1}\boldsymbol\pi$ has a pole of order $r$ at $s=\lambda$. This pole order is unchanged
by any representation giving the same signal. Scanning $\chi$ and measuring the coherence over time reveals the corresponding mode coalescence through its polynomial transient. 

Now we consider matched environments at the second order, i.e., they have the same stationary means $\mu=\langle\eta\rangle$ and two-point covariances $C_2(\tau)=\langle[\eta(t+\tau)-\mu][\eta(t)-\mu]\rangle$ for every temporal separation $\tau$. This ensures that their power spectra are identical. If such environments nevertheless produce different visible EP responses, some connected correlation above the second order must differ, because if all their multitime correlations were identical, they would give identical coherence at every $\chi$ and $t$. One must note that this implication does not require the noise to be sign symmetric. For several classical noise coordinates, our construction applies to $H(t)=H_0+\sum_a g_a\eta_a(t)V_a$ when $[H_0,V_a]=[V_a,V_b]=0$ for all $a,b$. In their common eigenbasis, one can express $\chi_a=g_a(v_{a,m}-v_{a,n})$ and $W-i\sum_a\chi_aX_a$. Second-order matching then requires equality of every mean $\mu_a$ and every auto- and cross-covariance $C_{ab}(\tau)=\langle[\eta_a(t+\tau)-\mu_a]
[\eta_b(t)-\mu_b]\rangle$ at all time separations.

We now construct a family of noise models for fitting experimental coherence decays by combining identical independent sources, motivated by descriptions of environments as collections of fluctuators \cite{Schriefl_2006,Bergli_2009,Ramon_2015}. Any stationary mean contributes only a deterministic phase, which we absorb into $H_0$. We therefore work with fluctuations whose average is zero. Now consider $M$ independent noise sources $\eta_1(t), \eta_2(t), \cdots, \eta_M(t)$, each following the same switching rules as the process $\eta(t)$, with each source initially in its stationary state. Their normalized sum, divided by $\sqrt M$, has the same variance and two-point correlation as a single source, and therefore the same power spectrum. We denote its connected $n$-time correlation by $C_n^{(M)}$, which is the part not built from lower-order correlations, and we suppress the time arguments for notational clarity. Now suppose that at a real coupling $\chi_{\rm EP}$, the tilted generator for a single source has a largest Jordan block of size $r$, whose $t^{r-1}$ term appears in the measured coherence. The normalized $M$-source noise then obeys
\begin{equation}
\begin{aligned}
\xi_M(t)&=\frac1{\sqrt M}\sum_{j=1}^M\eta_j(t),
& C_n^{(M)}&=M^{1-n/2}C_n^{(1)},\\
R_M&=1+M(r-1),
&|\chi_{\rm EP}^{(M)}|
&=\sqrt M |\chi_{\rm EP}|,
\end{aligned}
\label{copy-construction}
\end{equation}
where $R_M$ is the largest Jordan-block size of the combined generator at the coupling value $|\chi_{\rm EP}^{(M)}|$, and its highest time power also survives in the measured coherence (End Matter). Each noise source experiences the coupling $\chi/\sqrt M$, so all reach their EP at $|\chi_{\rm EP}^{(M)}|=\sqrt{M}|\chi_{\rm EP}|$. Their coherence factors multiply and taking the product of $M$ terms proportional to $t^{r-1}e^{\lambda t}$ gives a term proportional to $t^{M(r-1)}e^{M\lambda t}$, corresponding to the combined EP order $1+M(r-1)$ \cite{Prosen_2010,Wiersig_2025}. 

Suppose an experiment reveals a maximal visible EP of order $R_M$, say, through fitting a Jordan time transient through a polynomial of degree $R_M-1$ \cite{Ashida_2020}. Within the identical independent-source construction, the known single-source model then gives $M=(R_M-1)/(r-1)$ and predicts the scaling of the combined noise's higher-order connected correlations through Eq. (\ref{copy-construction}). As the number $M$ grows, the noise approaches Gaussian statistics at any fixed collection of times and the EP moves to stronger coupling values. This inference can be directly tested experimentally by fitting the measured coherence to $\Phi_M(\chi,t)=[\Phi_1(\chi/\sqrt M,t)]^M$ and comparing the predicted and observed EP parameters. For an unknown environment, choosing between switching models matters because they may share the same power spectrum yet predict different coherence decay as the quantum device is tuned. This raises the question: can an EP distinguish two switching environments even when their mean, power spectrum, and all three-time correlations are identical? This stronger matching is particularly relevant because third-order correlations can vanish identically even for non-Gaussian noise as a consequence of symmetry \cite{Ramon_2019}.

\textit{Matched switching environments.}
Let us for now move beyond combinations of identical sources and compare switching models with different transition rules. Their means, two-point correlations, and all three-time correlations agree, allowing us to ask whether their coherence decays still show different EP signatures. For definiteness we construct these models from a three-state switch whose measured noise takes the values $+a_\beta$, $0$, and $-a_\beta$ \cite{Bier_1996,Elston_1996}. Either of the nonzero states can move to zero at rate $\gamma$, while zero moves to each nonzero state at rate $\beta$. A visit to zero therefore lasts on average $1/(2\beta)$. In the stationary state, the two nonzero values each have probability $\beta/(\gamma+2\beta)$, and zero has
probability $\gamma/(\gamma+2\beta)$. We choose the convenient parametrization $a_\beta^2=\sigma^2(\gamma+2\beta)/(2\beta)$ so that every member of the $(\beta,\gamma)$ family has variance $\sigma^2$. Starting from noise value $x$, the expected value after delay $\tau$ is $xe^{-\gamma\tau}$, independent of $\beta$. With the Fourier convention $S_\beta(\omega)=\int_{-\infty}^{\infty}e^{i\omega\tau}C_{2,\beta}(\tau) d\tau$, this gives
\begin{equation}
C_{2,\beta}(\tau)=\sigma^2e^{-\gamma|\tau|},
 \quad \quad
S_\beta(\omega)
=\frac{2\sigma^2\gamma}{\gamma^2+\omega^2},
\label{matched-class}
\end{equation}
showing that the power spectrum stays fixed even though the switching rates and stationary probabilities differ. Interchanging the positive and negative states leaves the entire switching process unchanged while reversing every noise value, so all odd multitime correlations vanish. As we discuss below, the $\beta=\gamma/2$ and $\beta=\gamma$ cases exhibit EP$_3$ and EP$_2$ in their measured coherences, respectively. To locate the first statistical difference between these models, let $\eta_\beta$ denote the stationary noise. The fourth cumulant $\kappa_{4,\beta}=\langle\eta_\beta^4\rangle-3\sigma^4$ vanishes for Gaussian noise. For stationary $\eta_\beta$, the moment-generating function evaluates to $\langle e^{u\eta_\beta}\rangle=[\gamma+2\beta\cosh(u a_\beta)]/[\gamma+2\beta]$. Its logarithm generates the cumulants, with $u^4$ coefficient being $\kappa_{4,\beta}/4!$, i.e., 
\begin{equation}
\ln\langle e^{u\eta_\beta}\rangle=\frac{\sigma^2u^2}{2}
+\frac{\sigma^4(\gamma-4\beta)}{48\beta}u^4
+O(u^6),
\label{fourth-cumulant-beta}
\end{equation}
so $\kappa_{4,\beta}=\sigma^4(\gamma-4\beta)/(2\beta)$. We now compare two noise profiles of this family, corresponding to $\beta=\gamma/2$ and $\beta=\gamma$. For $t_1\leq t_2\leq t_3\leq t_4$, let $d_j=t_{j+1}-t_j$. For these noises, the connected four-time correlation is obtained by subtracting the three products of two-point correlations from the four-time moment \cite{Ramon_2019}. Averaging over the switching probabilities between consecutive observation times gives the difference
\begin{equation}
C_{4,c}^{(\gamma/2)}-C_{4,c}^{(\gamma)}
=\sigma^4
e^{-\gamma(d_1+d_3+2d_2)}
\left(1-\tfrac12e^{-\gamma d_2}\right)>0,
\label{fourth-correlation-difference}
\end{equation}
where $C_{4,c}^{(\beta)}$ denotes the connected four-time correlation of $\eta_\beta$, evaluated at $(t_1,t_2,t_3,t_4)$. At equal times the difference is $\sigma^4/2$. The matched environments therefore first differ in their four-time statistics.

\begin{figure}[t]
\centering
\begin{tikzpicture}
\pgfmathsetmacro{\xthree}{1/sqrt(2)}
\pgfmathsetmacro{\xtwo}{sqrt(2*0.418587820/3)}

\pgfmathdeclarefunction{matchedroot}{2}{%
  \pgfmathparse{%
    -4/3+2*sqrt(7/9-(#1)^2/2)*
    cos(acos(max(-1,min(1,-(20/27+(#1)^2)/
       (2*(7/9-(#1)^2/2)^(3/2)))))/3-120*(#2))%
  }%
}

\begin{groupplot}[
  group style={group size=1 by 2, vertical sep=1.20cm},
  myaxis,
  width=0.90\columnwidth,
  height=0.54\columnwidth,
  xmin=0, xmax=0.78,
  ymin=-3.1, ymax=0.12,
  xtick={0,0.2,0.4,0.6},
  ytick={-3,-2,-1,0},
  samples=280,
  no markers
]

\nextgroupplot[
  title={(a) $\beta=\gamma/2$},
  ylabel={$\lambda/\gamma$}
]
\addplot[very thick,deepteal,domain=0:\xthree]
  {-1+sqrt(max(0,1-2*x^2))};
\addplot[very thick,burntorange,domain=0:\xthree]
  {-1};
\addplot[very thick,deepmaroon,domain=0:\xthree]
  {-1-sqrt(max(0,1-2*x^2))};
\draw[dashed,black]
  (axis cs:\xthree,-3.1)--(axis cs:\xthree,0.12);
\addplot[only marks,mark=*,mark size=1.7pt,black]
  coordinates {(\xthree,-1)};
\node[anchor=west,font=\small]
  at (axis cs:0.42,-0.65) {EP$_3$};

\nextgroupplot[
  title={(b) $\beta=\gamma$},
  ylabel={$\lambda/\gamma$},
  xlabel={$\sigma|\chi|/\gamma$}
]
\addplot[very thick,deepteal,domain=0:\xtwo]
  {matchedroot(x,0)};
\addplot[very thick,burntorange,domain=0:\xtwo]
  {matchedroot(x,1)};
\addplot[very thick,deepmaroon,domain=0:\xtwo]
  {matchedroot(x,2)};
\draw[dashed,black]
  (axis cs:\xtwo,-3.1)--(axis cs:\xtwo,0.12);
\addplot[only marks,mark=*,mark size=1.7pt,black]
  coordinates {(\xtwo,-0.534428768)};
\node[anchor=west,font=\small]
  at (axis cs:0.55,-0.92) {EP$_2$};

\end{groupplot}
\end{tikzpicture}
\caption{\justifying{Real eigenvalues of the tilted three-state generators for the two switching rules with matched statistics up to the third order, as a function of the rescaled noise coupling $\sigma|\chi|/\gamma$. In (a), all the three branches meet at an EP$_3$. In (b), only two branches meet at an EP$_2$.}}
\label{fig2}
\end{figure}
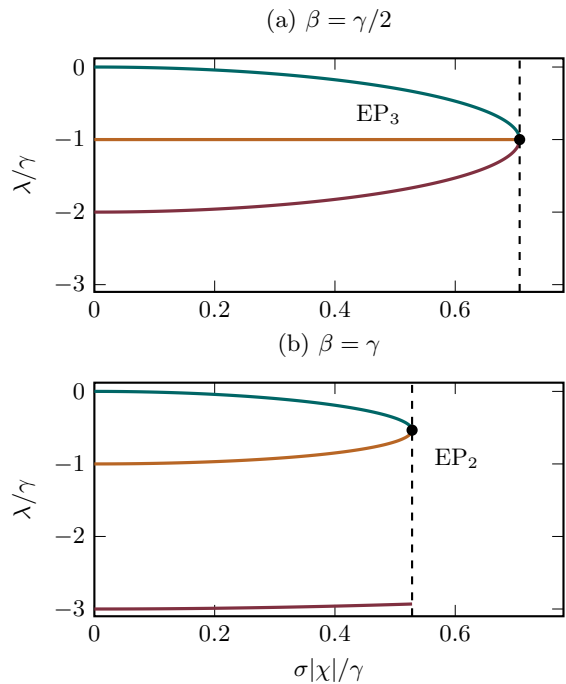

The coherence dynamics distinguishes these two environments with the following EP-coupling thresholds:
\begin{equation}
\begin{aligned}
\beta=\gamma/2\quad \implies \quad&
\mathrm{EP}_3,\quad \quad
|\chi_3|=\frac{\gamma}{\sqrt2 \sigma},\\
\beta=\gamma\quad \implies \quad&
\mathrm{EP}_2,\quad \quad
|\chi_2|\simeq0.52826 \frac{\gamma}{\sigma}.
\end{aligned}
\label{matched-EP-pair}
\end{equation}
Fig. \ref{fig2} shows the mode coalescences. To obtain this result (see End Matter), we start with the stationary probabilities, follow the phase accumulated as the noise switches among its three states, and sum over the final states to recover the device's coherence. The positive and negative states accumulate opposite noise-induced phases, while the zero state adds none. Changing $\beta$ changes how these contributions exchange weight, even after $a_\beta$ is adjusted to keep the power spectrum fixed. For $\beta = \gamma/2$ and the coupling value $|\chi_3|=\gamma/(\sqrt{2}\sigma)$ [Eq. (\ref{matched-EP-pair})], all the three decay modes merge and the measured coherence is $\Phi_{\gamma/2}(\chi_3,t)=e^{-\gamma t}(1+\gamma t/2)^2$. In contrast, for $\beta = \gamma$ and at the coupling value $|\chi_2|\simeq0.52826\gamma/\sigma$ [Eq. (\ref{matched-EP-pair})], two modes merge while the third remains distinct, and the coherence dynamics consequently has the form $\Phi_\gamma(\chi_2,t)=(A+Bt)e^{\lambda_{\rm EP} t}+Ce^{\lambda_3t}$, where $B\neq0$, $\lambda_{\rm EP}\simeq-0.53443\gamma$, and $\lambda_3\simeq-2.93114\gamma$. The quadratic (EP$_3$) and linear (EP$_2$) time factors provide a direct distinction between environments whose noise correlations agree through the third order. Notably, the EP$_3$/EP$_2$ distinction in Eq. (\ref{matched-EP-pair}) persists even for asymmetric noises with the same power spectrum containing two decay rates and equal nonzero three-time correlations. This follows from the fact that adding the same independent two-state process to each model leaves their fourth-order difference intact and multiplies both the coherence signals by a common factor, while the added process has no EP at either threshold \cite{supp}.

\begin{figure}[t]
\centering
\begin{tikzpicture}
\definecolor{cumslate}{RGB}{43,64,79}
\definecolor{cumbronze}{RGB}{155,115,76}
\begin{axis}[
  myaxis,
  width=0.90\columnwidth,
  height=0.58\columnwidth,
  xmin=0.90, xmax=3.57,
  ymin=-0.03, ymax=1.12,
  xtick={1,1.5,2,2.5,3,3.5},
  ytick={0,0.25,0.5,0.75,1},
  xlabel={$x=|\chi_{\rm EP}^{(M)}|/|\chi_{\rm EP}|$},
  ylabel={$|\kappa_n^{(M)}/\kappa_n^{(1)}|$},
  legend style={
    at={(0.97,0.97)},
    anchor=north east,
    draw=none,
    fill=none,
    font=\scriptsize
  },
  legend cell align=left
]

\addplot[very thick,cumslate,domain=1:sqrt(12),samples=150]
  {1/x^2};
\addlegendentry{$n=4$}

\addplot[very thick,cumbronze,domain=1:sqrt(12),samples=150]
  {1/x^4};
\addlegendentry{$n=6$}

\addplot[
  only marks,cumslate,mark=*,mark size=1.55pt,
  domain=1:12,samples=12
] ({sqrt(x)},{1/x});

\addplot[
  only marks,cumbronze,mark=square*,mark size=1.35pt,
  domain=1:12,samples=12
] ({sqrt(x)},{1/x^2});

\node[anchor=west,font=\scriptsize]
  at (axis cs:1.08,1.03) {EP$_2$};
\node[anchor=south west,font=\scriptsize]
  at (axis cs:1.44,0.53) {EP$_3$};
\node[anchor=south west,font=\scriptsize]
  at (axis cs:1.75,0.35) {EP$_4$};

\end{axis}
\end{tikzpicture}
\caption{\justifying{Scaling of cumulants with EP$_{\rm M+1}$ location at $x=|\chi_{\rm EP}^{(M)}|/|\chi_{\rm EP}|=\sqrt{M}$ for the symmetric telegraph hierarchy. Relative to a single source, the magnitudes of the fourth (slate blue, circles) and sixth (bronze, squares) equal-time cumulants fall as $M^{-1}=x^{-2}$ and $M^{-2}=x^{-4}$. The markers denote integer values of $M$. }}
\label{fig3}
\end{figure}
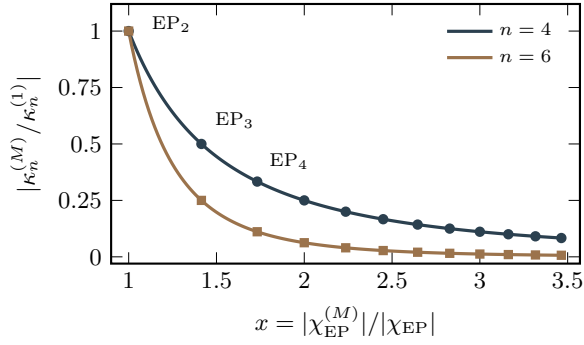

\begin{figure}[t]
\centering

\begin{tikzpicture}
\node[inner sep=0pt] (schematic) {
    \includegraphics[width=0.85\columnwidth]{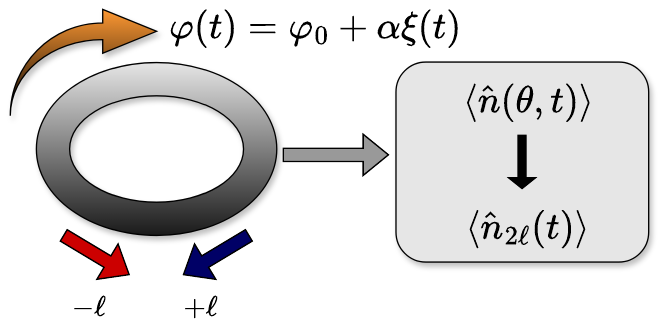}
};
\node[
    anchor=north,
    font=\bfseries
] at ([yshift=-1mm]schematic.south) {(a)};
\end{tikzpicture}

\vspace{1.5ex}

\begin{tikzpicture}
\begin{axis}[
  myaxis,
  width=0.90\columnwidth,
  height=0.58\columnwidth,
  xlabel={$\tau=\gamma t$},
  ylabel={
  $e^{-\lambda_{\rm EP}^{(\beta)}t}\frac{
  \overline{\langle\hat n_{2\ell}(t)\rangle}
  }{\langle\hat n_{2\ell}(t)\rangle_0}$
},
  xmin=0, xmax=5,
  ymin=0, ymax=13,
  xtick={0,1,2,3,4,5},
  ytick={0,2,4,6,8,10,12},
  domain=0:5,
  samples=250,
  legend style={
    at={(0.03,0.97)},
    anchor=north west,
    draw=none,
    fill=none,
    font=\scriptsize
  },
  legend cell align=left
]
\addplot[very thick,deepteal]
  {(1+x/2)^2};
\addlegendentry{$\beta=\gamma/2$ (EP$_3$)}

\addplot[very thick,burntorange]
  {0.998858758474701+0.537163997316937*x
   +0.001141241525299*exp(-2.396713695630304*x)};
\addlegendentry{$\beta=\gamma$ (EP$_2$)}

\end{axis}

\node[anchor=north,font=\bfseries]
  at ([yshift=-1mm]current bounding box.south) {(b)};
\end{tikzpicture}

\caption{\justifying{(a) Schematic of a ring condensate with opposite winding components and fluctuating synthetic flux. (b) Exponentially weighted normalized density harmonic at the respective EPs of the input noises $\eta_{\gamma/2}$ and $\eta_\gamma$, whose correlations agree up to the third order. The curves use separate $\alpha$ values corresponding to $\sigma|\chi_{2\ell}|/\gamma=1/\sqrt2$ and $0.52826$, respectively. The exponential weight uses coalesced eigenvalues $-\gamma$ and $-0.53443\gamma$, revealing quadratic EP$_3$ and nearly linear EP$_2$ time dependences.}}
\label{fig4}
\end{figure}

An intriguing observation is now in order. Within the three-state family, the EP$_3$ occurs only at $\beta/\gamma=1/2$. At this ratio, the noise is exactly the normalized sum of two identical, independent symmetric telegraph signals. This construction extends to $M$ such signals: $\xi_M(t)=\sigma M^{-1/2}\sum_{j=1}^{M}s_j(t)$, where each stationary $s_j=\pm1$ switches at the rate $\gamma/2$ \cite{Machlup_1954}. Cross-correlations between different signals vanish and the $M$ remaining terms cancel the normalization. Thus every member has the same two-point correlation $C_2(\tau)=\sigma^2e^{-\gamma|\tau|}$. Each source experiences the reduced effective coupling $\chi/\sqrt{M}$, so their elementary EPs occur together at $|\chi|=\gamma\sqrt{M}/(2\sigma)$. The independence property also makes their coherence factors multiply. At this coupling one finds
\begin{equation}
\Phi_M(\chi_{\rm EP}^{(M)},t)
=e^{-M\gamma t/2}
\left(1+\frac{\gamma t}{2}\right)^M ,
\label{telegraph-hierarchy}
\end{equation}
where the nonzero $t^M$ term establishes a coherence-visible EP$_{M+1}$ in the exact model. This simultaneous coalescence relies on identical elementary sources, since unequal switching rates or noise amplitudes would generally separate their EP couplings. The accessible eigenvalues are $\lambda_k=-M\gamma/2+(M-2k)\Delta_M$, where $\Delta_M=\sqrt{\gamma^2/4-\chi^2\sigma^2/M}$ and $k=0,\ldots,M$. Thus varying $\chi$ produces square-root splitting even though the Jordan block has order $M+1$; here that order is identified by the polynomial transient. The logarithm of the stationary moment-generating function is of the form $M\ln\cosh(u\sigma/\sqrt M)$, giving the equal-time fourth and sixth cumulants $\kappa_4^{(M)}=-2\sigma^4/M$ and
$\kappa_6^{(M)}=16\sigma^6/M^2$. Fig. \ref{fig3} shows how both of these scale with the EP location. For this telegraph family, an observed order $R=M+1$ predicts $\kappa_4^{(M)}=-2\sigma^4/(R-1)$, which is information absent from the power spectrum. 

\textit{Atomtronic readout.} We now present an illustrative readout using a ring-trapped condensate to map the distinct EP transients onto spatial interference. Ring condensates support quantized circulation \cite{Wright_2013}, while optical techniques provide related control and sensing capabilities \cite{Wright_2008,Chen_2018,Kumar_2021}. We consider a tightly confined atomic ring \cite{Dalibard_2011,Cooper_2019,Amico_2022}, with the condensate initially prepared with coherent winding components $+\ell$ and $-\ell$ [Fig. \ref{fig4}(a)], for example
$\psi(\theta,0)\propto c_+e^{i\ell\theta}+c_-e^{-i\ell\theta}$ with $c_+c_-\neq0$, so their interference produces fringes at the angular harmonic $q=2\ell$ \cite{Wright_2008}. After preparation, the atoms evolve on a smooth quasi-one-dimensional ring under a spatially uniform fluctuating synthetic flux \cite{Dalibard_2011,Pecci_2021}. For $N$ atoms of mass $m$ on a ring of radius $R$, we take
\begin{equation}
H_{\rm ring}(t)
=\frac{1}{2mR^2}\sum_{j=1}^{N}
[-i\partial_{\theta_j}-\varphi_0-\alpha\xi(t)]^2
+H_{\rm int},
\label{ring-model}
\end{equation}
where $\varphi_0$ is the steady flux, $\alpha$ controls its fluctuation, and $H_{\rm int}$ is any additional rotationally invariant contribution. The noise couples to the total angular momentum $L$ and other noise-dependent terms are proportional to the fixed particle number and contribute only a common phase. We monitor the density harmonic $\hat{n}_q=\int_0^{2\pi}d\theta e^{iq\theta}\hat n(\theta)$ \cite{Kumar_2016,Frometa-Fernandez_2026}. Since $[L,\hat n_q]=q\hat n_q$, each noise history multiplies it by a phase. Averaging this complex component over repeated images gives
\begin{equation}
\frac{\overline{\langle\hat n_q(t)\rangle}}
     {\langle\hat n_q(t)\rangle_0}
=\Phi(\chi_q,t),
 \quad \quad
\chi_q=\frac{2\alpha q}{2mR^2},
\label{ring-fringe}
\end{equation}
where $\langle\hat n_q(t)\rangle_0$ is the noise-free expectation at time $t$ and we assume $\langle\hat{n}_q(t)\rangle_0\neq0$. Eq. (\ref{ring-fringe}) holds for any classical flux process in the ideal ring model, including in the presence of rotationally invariant interactions \cite{supp}. At any common $\alpha$, the two flux noises have identical correlations through the third order. Scanning $\alpha$ locates their distinct EP thresholds, shown in Fig. \ref{fig4}(b) at separately tuned settings. The plot weights each normalized density harmonic by $e^{-\lambda_{\rm EP}^{(\beta)}t}$, where $\lambda_{\rm EP}^{(\beta)}$ is its coalesced eigenvalue. This gives $(1+\gamma t/2)^2$ for $\beta=\gamma/2$ and $A+Bt+C e^{(\lambda_3-\lambda_{\rm EP})t}$ for $\beta=\gamma$, with $C\simeq0.00114$, displaying distinct coherence-visible EP orders.

\textit{Discussion.} In this work we have shown the importance of exceptional points in probing non-Gaussian noise of direct relevance to real experimental platforms. Specifically, we have shown that noise environments with identical means and identical two- and three-time correlation functions for all time arguments can produce different EP orders in a quantum probe. In particular, our matched pair first differs at the fourth order, while the noise-averaged probe response exhibits an EP$_3$ in one case and an EP$_2$ in the other. We have also shown that for identical independent sources of a known type, the EP order of the normalized multi-source response fixes how its higher connected correlations scale, while the power spectrum stays constant, provided the maximal Jordan block at the single-source EP is visible in its measured decay. The realizable atomtronic model maps the EP$_3$/EP$_2$ distinction onto a density harmonic, providing a spatial readout. This work therefore shows that different noise sources, while in agreement at the level of the lower-order statistics, can produce distinct EP signatures in the dephasing response of quantum devices. Our results open new avenues for experimental noise characterization, including benchmarking EP-based signatures against higher-order spectroscopy with coherence measurements in platforms like superconducting qubits \cite{Bylander_2011,Sung_2019}, trapped ions \cite{Kotler_2013}, and nuclear-spin systems \cite{Chen-noise_2018}.\\

\noindent
\textit{Acknowledgements.} A.G. thanks Sushanta Dattagupta for inspiring discussions. A.G. and M.B. thank the Air Force Office of Scientific Research (FA9550-23-1-0259) for support.

\clearpage
\begin{center}
\textbf{\large End Matter}
\end{center}

Here we supply the proof of the independent-copy scaling summarized in Eq. (\ref{copy-construction}), the derivations underlying the matched-noise construction and its EPs in Eqs. (\ref{matched-class})-(\ref{matched-EP-pair}), and also the derivation behind the symmetric-telegraph hierarchy used in Eq. (\ref{telegraph-hierarchy}), including the coalescing spectrum, the visible EP$_{M+1}$ structure, and the associated scaling of the higher noise statistics.

\setcounter{equation}{0}
\renewcommand{\theequation}{A\arabic{equation}}
\subsection{Independent-copy analysis}
\label{sec:end-copy-theorem}

Let $G_1(\chi)=W-i\chi X$ be the tilted generator of one centered
stationary source with stationary vector $\boldsymbol\pi$. For $M$
independently initialized copies, the joint preparation is $\boldsymbol\pi^{\otimes M}$ and the coherence readout is
$(\mathbf{1}^T)^{\otimes M}$. The normalized sum couples each factor
with strength $\chi/\sqrt M$, so its tilted generator is a sum of
operators acting on different tensor factors. These operators act on different sources, so they commute and the exponential factorizes. Averaging over the sources' stationary initial
states and summing over all their final states then gives
\begin{equation}
\begin{aligned}
G_M(\chi)
&=
\sum_{j=1}^{M}
I^{\otimes(j-1)}
\otimes
G_1\!\left(\frac{\chi}{\sqrt M}\right)
\otimes
I^{\otimes(M-j)},\\
\Phi_M(\chi,t)
&=
\left[
\Phi_1\!\left(\frac{\chi}{\sqrt M},t\right)
\right]^M .
\end{aligned}
\label{end-copy-factorization}
\end{equation}
This is the standard Kronecker-sum composition of independent non-Hermitian factors \cite{Prosen_2010,Wiersig_2025}. An EP of the single-source tilted generator at $\chi_{\rm EP}$ is therefore inherited at $\chi=\sqrt M \chi_{\rm EP}$ by the composition of $M$ identical noise sources.

Let $J=\lambda I+N$ be a largest single-source Jordan block of size $r$, where $N$ is nilpotent, i.e., $N^r=0$ but $N^{r-1}\neq0$. On the tensor product of $M$ such blocks we define $N^{(j)}=I^{\otimes(j-1)}\otimes N\otimes I^{\otimes(M-j)}$ and $d=M(r-1)$. Since the $N^{(j)}$ commute, every multinomial term of total degree $d+1$ contains some $N^r$ and vanishes, whereas at degree $d$ the only surviving exponent assignment is $r-1$ on every factor, given by
\begin{equation}
\begin{aligned}
&\left(\sum_jN^{(j)}\right)^{d+1}=0,\\
&\left(\sum_jN^{(j)}\right)^d
= \frac{d!}{[(r-1)!]^M}
\left(N^{r-1}\right)^{\otimes M}\neq0 .
\end{aligned}
\label{end-copy-jordan}
\end{equation}
The largest block therefore has size $d+1=1+M(r-1)$.
More generally, combining blocks of sizes $r_1,\ldots,r_M$ produces no block larger than $1+\sum_j(r_j-1)$ \cite{Prosen_2010,Wiersig_2025}.

We next prove that it is visible in the time dynamics. At the EP of a single source we write \cite{Ashida_2020}
\begin{equation}
\Phi_1(\chi_{\rm EP},t)
=
\sum_\lambda e^{\lambda t}
\sum_{p=0}^{r_\lambda-1}c_{\lambda,p}t^p ,
\label{end-seed-jordan-expansion}
\end{equation}
where $r_\lambda$ is the largest Jordan-block size at eigenvalue $\lambda$. Let us collect the degree-$(r-1)$ terms as $t^{r-1}F(t)$, where $F(t)=\sum_{\lambda:r_\lambda=r}c_{\lambda,r-1}e^{\lambda t}$. Visibility implies $F\not\equiv0$, since distinct exponentials are linearly independent. In $\Phi_1^M$, the highest possible degree $d=M(r-1)$ requires taking degree $r-1$ from every factor. Its contribution is therefore $t^dF(t)^M$. Since $F(t)^M$ is a
nonzero finite sum of exponentials, at least one exponential retains degree $d$ after grouping equal exponents, which lower-degree terms cannot cancel, proving the visibility of a maximal combined block. 

Finally, the connected-correlation scaling of Eq. (\ref{copy-construction}) follows from the product structure. We define the cumulant-generating functional $\mathcal{K}[f]=\ln\langle e^{i\int dtf(t)\eta(t)}\rangle$ \cite{Gardiner_2009}. For the normalized sum, the exponential factorizes into $M$ terms with functional argument $f/\sqrt M$. Independence of the sources therefore gives
\begin{equation}
\mathcal{K}_M[f]
=
M\mathcal{K}_1[f/\sqrt M].
\label{end-cumulant-functional-scaling}
\end{equation}
Each of $n$ functional derivatives contributes a factor $M^{-1/2}$, while the prefactor contributes $M$, yielding exactly the factor $M^{1-n/2}$ in Eq. (\ref{copy-construction}).

\setcounter{equation}{0}
\renewcommand{\theequation}{B\arabic{equation}}
\subsection{Exact matched pair and visible EPs}
\label{sec:end-matched-pair}

For states ordered as $+,0,-$, the three-state family used in the
main text is
\begin{equation}
\begin{aligned}
W_\beta&=
\begin{pmatrix}
-\gamma&\beta&0\\
\gamma&-2\beta&\gamma\\
0&\beta&-\gamma
\end{pmatrix},
&
\mathbf{x}_\beta&=a_\beta(1,0,-1)^T,\\
\boldsymbol\pi_\beta&=
\frac{(\beta,\gamma,\beta)^T}{\gamma+2\beta},
&
a_\beta^2&=
\frac{\sigma^2(\gamma+2\beta)}{2\beta}.
\end{aligned}
\label{end-pair}
\end{equation}
With $X_\beta=\operatorname{diag}(\mathbf{x}_\beta)$, the explicit
expressions in Eq. (\ref{end-pair}) satisfy $W_\beta\boldsymbol\pi_\beta=0$, $\mathbf{x}_\beta^T\boldsymbol\pi_\beta=0$, $W_\beta^T\mathbf{x}_\beta=-\gamma\mathbf{x}_\beta$, and $\mathbf{x}_\beta^TX_\beta\boldsymbol\pi_\beta=\sigma^2$. For $\tau\geq0$, averaging the product of the two noise values over the stationary switching process gives $C_{2,\beta}(\tau)= \mathbf{x}_\beta^Te^{W_\beta\tau}X_\beta\boldsymbol\pi_\beta$ \cite{Kubo_1963}. Since $\mathbf{x}_\beta^Te^{W_\beta\tau}
=e^{-\gamma\tau}\mathbf{x}_\beta^T$, the relation $W_\beta^T\mathbf{x}_\beta=-\gamma\mathbf{x}_\beta$ immediately yields the common covariance in Eq. (\ref{matched-class}) and stationarity extends it to negative $\tau$. Detailed higher-order correlations are derived in \cite{supp}.

To locate the EPs, we consider Laplace transformation of Eq. (\ref{coherence-factor}), giving $\widehat\Phi_\beta(s) = N_\beta(s)/D_\beta(s)$, with
\begin{equation}
\begin{aligned}
D_\beta(s)
&=s(s+\gamma)(s+\gamma+2\beta)+y^2(s+2\beta),\\
N_\beta(s)
&=(s+\gamma)(s+\gamma+2\beta)
+\frac{\gamma y^2}{\gamma+2\beta},
\end{aligned}
\label{end-pair-laplace}
\end{equation}
where $y=\chi a_\beta$. To determine when a triple root is possible, we set
$b=\beta/\gamma$, $u=s/\gamma$, and $z=y^2/\gamma^2$. Then
\begin{equation}
\frac{D_\beta(\gamma u)}{\gamma^3}
=u^3+2(1+b)u^2+(1+2b+z)u+2bz .
\label{end-triple-condition}
\end{equation}
Comparison with the triple-root polynomial $(u-u_0)^3$ gives
$u_0=-2(1+b)/3$, $z=(1+2b+4b^2)/3$, and
$(2b-1)(16b^2+11b+4)=0$. Since $b>0$, a triple root requires $b=1/2$, for which $z=1$. 

For $\beta=\gamma/2$, the first coupling in Eq. (\ref{matched-EP-pair}) gives $y=\gamma$, so $D_\beta=(s+\gamma)^3$, while $N_\beta(-\gamma)=\gamma^2/2\neq0$. The projected resolvent therefore has an uncancelled cubic pole. Its inverse Laplace transform is $\Phi_{\gamma/2}(\chi_3,t)=e^{-\gamma t}(1+\gamma t/2)^2$, proving the existence of EP$_3$.

For $\beta=\gamma$, we set $u=s/\gamma$ and $z=y^2/\gamma^2$. Eq. (\ref{end-triple-condition}) becomes $D_\gamma(\gamma u)/\gamma^3=u^3+4u^2+(3+z)u+2z$ and its discriminant $4(9-23z+4z^2-z^3)$ vanishes exactly when eigenvalues coincide. The polynomial in parentheses decreases strictly for $z\geq0$, since its derivative is $-23+8z-3z^2<0$. It starts positive and tends to $-\infty$, giving one positive zero $z_{\rm EP}=0.418587820\ldots$.
The repeated root is $u_{\rm EP}=-0.534428768\ldots$, and the third root $-4-2u_{\rm EP}$ is distinct. Moreover, $N_\gamma(\gamma u_{\rm EP})/\gamma^2=2+4u_{\rm EP}/3>0$, so the double pole survives and the EP$_2$ is visible. Using $a_\gamma^2=3\sigma^2/2$ gives the second coupling in Eq. (\ref{matched-EP-pair}).

\setcounter{equation}{0}
\renewcommand{\theequation}{C\arabic{equation}}
\subsection{Exact symmetric-telegraph hierarchy}
\label{sec:end-telegraph-hierarchy}

For a stationary symmetric telegraph process $s(t)=\pm1$ whose two states interchange at rate $\gamma/2$, $\langle s(t+\tau)s(t)\rangle=e^{-\gamma|\tau|}$ \cite{Machlup_1954}. Each source has zero mean, so for distinct independent copies, one has $\langle s_j(t+\tau)s_k(t)\rangle=0$ when $j\neq k$. Thus, in the double sum defining the covariance of $\xi_M$, only the $M$ terms with $j=k$ survive and cancel the normalization $1/M$. This gives $C_2^{(M)}(\tau)=\sigma^2e^{-\gamma|\tau|}$, independently of $M$, matching the covariance in Eq. (\ref{matched-class}). For $M=2$, the normalized sum takes the three values $\sqrt2\sigma,0,-\sqrt2\sigma$. Either of the outer states reaches the middle state at a total rate $\gamma$, while the middle reaches either specified outer state at rate $\gamma/2$. This is exactly Eq. (\ref{end-pair}) with $\beta=\gamma/2$.

For general $M$, let $k$ count sources in state $-1$. Of the $M$ sources, $M-k$ can increase $k$ and $k$ can decrease it. The reduced birth-death process \cite{Karlin_1965}
therefore has
\begin{equation}
\begin{aligned}
k\to k+1:\frac{(M-k)\gamma}{2}, \quad \quad
k\to k-1:\frac{k\gamma}{2},\\
\pi_k=2^{-M}\binom{M}{k}, \quad \quad
x_k=\frac{\sigma(M-2k)}{\sqrt M}.
\end{aligned}
\label{end-telegraph-reduction}
\end{equation}
The stationary product state and the readout
$(\mathbf{1}^T)^{\otimes M}$ are permutation symmetric and the full tilted generator commutes with source permutations. Evolution from the stationary preparation therefore remains in the symmetric
subspace spanned by $k=0,\ldots,M$, whose dimension is $M+1$.

The spectrum is simplest before this reduction. Applying the stochastic-Liouville tilt of Eq. (\ref{coherence-factor}) \cite{Kubo_1963}, each constituent source, coupled with strength $\chi/\sqrt{M}$ and prepared in the stationary vector $\boldsymbol\pi_{\rm tel}=(1/2,1/2)^T$, has the tilted generator
\begin{equation}
G_{\rm tel}(\chi/\sqrt{M})=
\begin{pmatrix}
-\gamma/2-i\chi\sigma/\sqrt M&\gamma/2\\
\gamma/2&-\gamma/2+i\chi\sigma/\sqrt M
\end{pmatrix},
\label{end-single-telegraph-generator}
\end{equation}
The eigenvalues are $-\gamma/2\pm\Delta_M$, where $\Delta_M=\sqrt{\frac{\gamma^2}{4}-\frac{\chi^2\sigma^2}{M}}$. The full $M$-source tilted generator is $G_M(\chi)=\sum_{j=1}^{M}
I_2^{\otimes(j-1)}\otimes
G_{\mathrm{tel}}(\chi/\sqrt M)\otimes
I_2^{\otimes(M-j)}$, where $I_2$ is the $2\times2$ identity. Its eigenvalues are obtained by adding one eigenvalue from each constituent generator. If $k$ factors contribute the minus branch and $M-k$ the plus branch, the symmetric-sector eigenvalue is
\begin{equation}
\lambda_k
=
-\frac{M\gamma}{2}
+(M-2k)\Delta_M,
 \quad \quad k=0,\ldots,M.
\label{end-telegraph-spectrum}
\end{equation}
All the $M+1$ branches therefore meet at $\Delta_M=0$, giving the EP coupling quoted in the main text and the $\sqrt M$ displacement in Eq. (\ref{copy-construction}).

The coherence factor of an individual source is
$\phi_M(\chi,t)=\mathbf{1}^T e^{G_{\rm tel}(\chi/\sqrt{M})t} \boldsymbol\pi_{\rm tel}$, where the readout row is $\mathbf{1}^T=(1,1)$. Exponentiating Eq. (\ref{end-single-telegraph-generator}) gives
\begin{equation}
\frac{\phi_M(\chi,t)}{e^{-\gamma t/2}}
=
\left[
\cosh(\Delta_Mt)
+\frac{\gamma}{2\Delta_M}\sinh(\Delta_Mt)
\right].
\label{end-single-telegraph-coherence}
\end{equation}
The accumulated phase of $\xi_M$ is a sum over the $M$ sources, so its exponential is a product of $M$ single-source phase factors. For independent sources the expectation of that product equals the product of the expectations, that is
\begin{equation}
\frac{\Phi_M(\chi,t)}{e^{-M\gamma t/2}}
=
\left[
\cosh(\Delta_Mt)
+\frac{\gamma}{2\Delta_M}\sinh(\Delta_Mt)
\right]^M .
\label{end-telegraph-coherence}
\end{equation}
At $\Delta_M=0$, using $\sinh(\Delta_Mt)/\Delta_M\to t$ reduces this expression to Eq. (\ref{telegraph-hierarchy}). Its highest term is $t^Me^{-M\gamma t/2}$ with a nonzero coefficient. Since a size-$r$ Jordan block can generate powers only through $t^{r-1}$, this term
requires a visible block of size at least $M+1$. The accessible symmetric sector itself has dimension $M+1$, so the block has exactly that size and exhausts the sector at the EP. Because varying $\chi$ changes every identical source in the same way, the eigenvalue separations in Eq. (\ref{end-telegraph-spectrum}) retain the square-root dependence of each constituent, even though the combined EP has order $M+1$.

Since each sign is equally probable, one writes $\langle e^{v s_j}\rangle=\cosh v$, so we have the $\cosh$ expansion
\begin{equation}
\ln\langle e^{u\xi_M}\rangle =
\frac{\sigma^2u^2}{2}
-\frac{\sigma^4u^4}{12M}
+\frac{\sigma^6u^6}{45M^2}
+O(u^8).
\label{end-telegraph-cgf}
\end{equation}
All the multitime connected correlations obey the general scaling proved in Sec. \ref{sec:end-copy-theorem}. Thus this hierarchy realizes, within one explicit finite-state family, the fixed second-order statistics, growing visible EP order, and higher connected statistics summarized in Eq. (\ref{copy-construction}).

\clearpage

\setcounter{section}{0}
\setcounter{equation}{0}
\setcounter{figure}{0}
\setcounter{table}{0}
\setcounter{page}{1}

\renewcommand{\thesection}{S\arabic{section}}
\renewcommand{\theequation}{S\arabic{equation}}
\renewcommand{\thefigure}{S\arabic{figure}}
\renewcommand{\thetable}{S\arabic{table}}
\renewcommand{\thepage}{S\arabic{page}}

\onecolumngrid 
\begin{center}
    \textbf{\large Supplementary Material for:\\ Exceptional Points as Witnesses of Non-Gaussian Noise} \\
    \vspace{4mm}
    Aritra Ghosh$^1$, Subhashish Banerjee$^2$, and M. Bhattacharya$^1$,\\
    {\it $^1$School of Physics and Astronomy, Rochester Institute of Technology,\\ 84 Lomb Memorial Drive, Rochester, New York 14623, USA}\\
    {\it $^2$Department of Physics, Indian Institute of Technology Jodhpur, Rajasthan 342030, India}
\end{center}

\section{General results for stationary switching noise}
\label{sec:supp-general}

\subsection{Phase averaging and two-point correlations}
\label{sec:supp-conditional}

We now derive two formulas used throughout the paper. The first determines the phase measured by a dephasing probe for an arbitrary finite-state switching process. The second determines its two-point correlations directly from the same transition generator, allowing us to verify when different environments are identical at the second order.

Let $J_t$ be a stationary finite-state Markov chain with column generator $W$ and stationary vector $\boldsymbol\pi$, so that $\mathbf{1}^TW=0$, $W\boldsymbol\pi=0$, and $\mathbf{1}^T\boldsymbol\pi=1$. Its measured coordinates are $\xi_a(t)=x_{a,J_t}$, with
$X_a=\operatorname{diag}(\mathbf{x}_a)$. To resolve the phase average according to the final Markov state, we define $\phi_j(t)=\left\langle
\delta_{J_t,j}\exp[-i\int_0^t ds\sum_a f_a(s)\xi_a(s)]\right\rangle$, where $\delta_{J_t,j}$ selects trajectories ending in the state $j$. The standard Feynman-Kac construction for additive functionals of a Markov process \cite{Kubo_1963,Fitzsimmons_1999} then gives
\begin{equation}
\dot{\boldsymbol\phi}(t)
=
\left[
W-i\sum_a f_a(t)X_a
\right]\boldsymbol\phi(t),\quad \quad
\boldsymbol\phi(0)=\boldsymbol\pi,\quad \quad
\Phi[f]=\mathbf{1}^T\boldsymbol\phi(t),
\label{supp-tilt}
\end{equation}
where
$\Phi[f]=
\langle\exp[-i\int_0^t ds\sum_a f_a(s)\xi_a(s)]\rangle$.
The single-coordinate, constant-field case is
Eq. (\ref{coherence-factor}) of the main text. If $\rho_j(t)$ is the unnormalized probe state resolved by Markov state $j$, then
$\dot\rho_j=-i[H_0+g x_jV,\rho_j]+\sum_k W_{jk}\rho_k$, with $\rho_j(0)=\pi_j\rho(0)$. Its $(m,n)$ coherence evolves under the tilted generator $W-i\chi X$ after the removal of the ordinary phase from $H_0$. Thus the tilted generator is an invariant coherence block of the joint classical-quantum stochastic Liouvillian. Eq. (\ref{supp-tilt}) averages the phases generated by otherwise unitary realizations and does not describe the evolution generator of an individual realization.

We next derive the corresponding two-point correlations. Let $\mu_a=\mathbf{x}_a^T\boldsymbol\pi$. Weighting the probability of the state at the earlier time by $x_{b,J_t}$ produces the vector $X_b\boldsymbol\pi$. Averaging over the possible state changes during the interval $\tau\geq0$ then gives
\begin{equation}
\begin{aligned}
\langle\xi_a(t+\tau)\xi_b(t)\rangle
&=
\mathbf{x}_a^Te^{W\tau}X_b\boldsymbol\pi,\\
C_{ab}(\tau)
&=
\mathbf{x}_a^Te^{W\tau}X_b\boldsymbol\pi-\mu_a\mu_b,\\
C_{ab}(-\tau)&=C_{ba}(\tau).
\end{aligned}
\label{supp-covariance}
\end{equation}
Thus, matching the mean of each noise coordinate and the auto- and cross-covariances at every time separation gives identical second-order noise statistics for a probe with the same linear couplings. The corresponding spectral matrix is the two-sided Fourier transform of $C_{ab}(\tau)$.

\subsection{Higher correlations of the matched three-state family}
\label{sec:supp-pair-correlations}

In this subsection we identify explicitly where the matched three-state environments begin to differ. We first show that all their odd correlations vanish, then calculate the equal-time fourth cumulant and the full four-time correlations used in the main text. These results demonstrate directly that fixing every two-point correlation does not fix the higher switching statistics.

The family in Eq. (\ref{end-pair}) is invariant under interchange
of its two outer states. If $P$ performs this interchange, then
\begin{equation}
PW_\beta P=W_\beta, \quad \quad
PX_\beta P=-X_\beta, \quad \quad
P\boldsymbol\pi_\beta=\boldsymbol\pi_\beta.
\label{supp-pair-parity}
\end{equation}
Exchanging the two outer states using $P$ leaves the switching probabilities and stationary distribution unchanged, but reverses the sign of each noise value in the multitime product. Every odd multitime moment therefore equals its own negative and therefore vanishes.

The equal-time fourth moment follows directly from the stationary
weights as
\begin{equation}
\langle\eta_\beta^4\rangle =
\frac{2\beta}{\gamma+2\beta}a_\beta^4 =
\sigma^4\frac{\gamma+2\beta}{2\beta},
\label{supp-pair-fourth-moment}
\end{equation}
and since $\langle\eta_\beta^2\rangle=\sigma^2$, subtracting $3\sigma^4$ gives the fourth cumulant used in Eq. (\ref{fourth-cumulant-beta}) of the main text.

We now consider noise values at four observation times. For $t_1\leq t_2\leq t_3\leq t_4$ and $d_j=t_{j+1}-t_j$, averaging their product using the stationary initial probabilities and the switching probabilities between consecutive observations gives
\begin{equation}
M_{4,\beta}
=
\mathbf{1}^T
X_\beta e^{W_\beta d_3}
X_\beta e^{W_\beta d_2}
X_\beta e^{W_\beta d_1}
X_\beta\boldsymbol\pi_\beta .
\label{supp-fourth-propagation}
\end{equation}
Subtracting the three products of two-point correlations from Eq. (\ref{supp-fourth-propagation}) gives, for any $\beta>0$, the expression
\begin{equation}
C_{4,c}^{(\beta)}
=\sigma^4e^{-\gamma(d_1+d_3)}
\left[
\frac{\gamma}{2\beta}e^{-(\gamma+2\beta)d_2}
-2e^{-2\gamma d_2}
\right].
\label{supp-fourth-members}
\end{equation}
Setting $\beta=\gamma/2$ and $\beta=\gamma$ and subtracting reproduces Eq. (\ref{fourth-correlation-difference}) of the main text. The difference also affects ordinary coherence measurements. At fixed $t$, expansion about $\chi=0$ gives
\begin{equation}
\ln\Phi_{\gamma/2}(\chi,t)-\ln\Phi_\gamma(\chi,t) =
\frac{\chi^4}{4!}
\int_{[0,t]^4}\!dt_1 dt_2 dt_3 dt_4 
\left[
C_{4,c}^{(\gamma/2)}-C_{4,c}^{(\gamma)}
\right]
+O(\chi^6),
\label{supp-weak-coupling-difference}
\end{equation}
where the correlators in the integrand carry the displayed time arguments. The common second-order term cancels and odd orders vanish by outer-state symmetry. Since the fourth-order difference is positive, the coherences differ at sufficiently small nonzero coupling. Their distinct EP orders, established in the End Matter, provide an additional signature of this higher-order statistical difference.

Finally let us argue that an equal-time cumulant is insufficient to conclude Gaussianity, motivating our use of unequal-time correlations in this work. For the case $\beta=\gamma/4$, Eq. (\ref{supp-fourth-members}) gives
\begin{equation}
C_{4,c}^{(\gamma/4)}
=
2\sigma^4
\left(e^{\gamma d_2/2}-1\right)
e^{-\gamma(d_1+2d_2+d_3)} .
\label{supp-fourth-quarter}
\end{equation}
It vanishes at equal times, and more generally whenever $d_2=0$, consistently with $\kappa_{4,\gamma/4}=0$, but is nonzero whenever $d_2>0$. Thus the separated-time fourth-order statistics can remain non-Gaussian even when the equal-time fourth cumulant vanishes.

\subsection{Common asymmetric background}
\label{sec:supp-background}

We now test whether the EP distinction depends on the special symmetry of the three-state examples. We add a fixed asymmetric two-state process to every environment and show that the resulting noises still agree through the third order, while their fourth-order distinction and visible EP orders remain unchanged. The construction also introduces a second decay rate into the two-point correlation.

Let $b(t)$ be an independent stationary two-state background process taking the values $b_+=\sqrt2\nu$ and $b_-=-\nu/\sqrt2$, where $\nu>0$ is its stationary standard deviation. In the state order $(+,-)$, its transition generator, noise-value vector $\mathbf{b}=(b_+,b_-)^T$, and stationary distribution are
\begin{equation}
W_b=
\begin{pmatrix}
-2\gamma&\gamma\\
2\gamma&-\gamma
\end{pmatrix},
 \quad \quad
\mathbf{b}=
\begin{pmatrix}
\sqrt2 \nu\\
-\nu/\sqrt2
\end{pmatrix},
 \quad \quad
\boldsymbol\pi_b=
\begin{pmatrix}
1/3\\
2/3
\end{pmatrix}.
\label{supp-background-definition}
\end{equation}
One can easily verify that
\begin{equation}
W_b^T\mathbf{b}=-3\gamma\mathbf{b}, \quad \quad
\langle b\rangle=0, \quad \quad
\langle b^2\rangle=\nu^2, \quad \quad
\langle b^3\rangle=\frac{\nu^3}{\sqrt2}, \quad \quad
\langle b^4\rangle=\frac{3\nu^4}{2},
\label{supp-background-moments}
\end{equation}
and Eq. (\ref{supp-covariance}) therefore gives $C_2^b(\tau)=\nu^2e^{-3\gamma|\tau|}$, while the last two moments give $\kappa_3(b)=\nu^3/\sqrt2$ and $\kappa_4(b)=-3\nu^4/2$.

The full three-time correlation function follows from the pointwise identity $b^2=\nu^2+(\nu/\sqrt2)b$. For ordered times $t_1\leq t_2\leq t_3$, following the two-state dynamics between consecutive observation times and using $W_b^T\mathbf{b}=-3\gamma\mathbf{b}$ gives
\begin{equation}
C_{3,c}^{b}(t_1,t_2,t_3)
=
\frac{\nu^3}{\sqrt2}
e^{-3\gamma(t_3-t_1)}.
\label{supp-background-third}
\end{equation}
The three-state processes have vanishing odd correlations, so adding
the same independent $b(t)$ to either one gives identical
three-time correlations for all time arguments.

Now for the two matched environments let us define $\zeta_\beta(t)=\eta_\beta(t)+b(t)$. Because the two components are independent, the logarithm of their joint characteristic function is the sum of the two logarithms, and the connected cumulants therefore add. In particular, one can write
\begin{equation}
\kappa_4(\zeta_{\gamma/2})
=-\sigma^4-\frac{3\nu^4}{2},\quad \quad
\kappa_4(\zeta_\gamma)
=-\frac{3\sigma^4}{2}-\frac{3\nu^4}{2}.
\label{supp-background-fourth}
\end{equation}
Their difference remains $\sigma^4/2$, equal to the equal-time limit of Eq. (\ref{fourth-correlation-difference}).

We now show that the added process does not remove the EP distinction. The discriminant of its tilted two-state generator is
\begin{equation}
\Delta_b^{\rm disc}(\chi)
=
9\gamma^2
-\chi^2(b_+-b_-)^2
+2i\gamma\chi(b_+-b_-).
\label{supp-background-discriminant}
\end{equation}
For real $\chi\neq0$ and $\nu\neq0$, the imaginary part is nonzero, so the two background eigenvalues are distinct. Its Jordan blocks are therefore one dimensional at every relevant real coupling. In the Kronecker sum with the three-state generator, each non-diagonalizable block is consequently only shifted by one of these background eigenvalues and retains its size \cite{Prosen_2010,Wiersig_2025}. Independence also gives $\Phi_{\zeta_\beta}(\chi,t)=\Phi_\beta(\chi,t)\Phi_b(\chi,t)$.
Because the background has two distinct eigenvalues, $\Phi_b=\sum_{\rho=\pm}d_\rho e^{\mu_\rho t}$ with $\mu_+\neq\mu_-$, and at least one $d_\rho$ is nonzero since $\Phi_b(\chi,0)=1$. At each EP considered here, the highest-degree polynomial term in $\Phi_\beta$ has a single seed exponent. Multiplication by any nonzero background term therefore preserves that polynomial degree at a shifted exponent, and the distinct background exponents cannot cancel it. The visible EP$_3$/EP$_2$ distinction consequently survives the common asymmetric background.

\subsection{Gaussian limit of normalized independent sums}
\label{sec:supp-gaussian}

Here we show that any fixed collection of noise values becomes Gaussian, that the accumulated phase measured by the probe has the corresponding Gaussian limit, and derive the Ornstein-Uhlenbeck coherence factor for an exponential covariance.

Consider the centered independent-copy construction of Eq. (\ref{copy-construction}). We denote the covariance of
one centered source by $C(\tau)=\langle\eta(t+\tau)\eta(t)\rangle$. For fixed times $t_1,\ldots,t_k$, the random vectors $(\eta_j(t_1),\ldots,\eta_j(t_k))$ are independent and identically distributed across $j$ and are bounded. The multivariate central-limit theorem therefore gives a Gaussian limit for
$(\xi_M(t_1),\ldots,\xi_M(t_k))$ as $M\to\infty$. This agrees with
the connected-correlation scaling in the End Matter.

The accumulated noise entering the coherence can be treated with the
ordinary central-limit theorem. For fixed $t$, we define $Y_j(t)=\int_0^t ds \eta_j(s)$. Then
$\int_0^t ds\xi_M(s)=M^{-1/2}\sum_jY_j(t)$. The $Y_j(t)$ are independent and identically distributed across copies, and their variance is $\operatorname{Var}Y_j(t)=
\int_0^t ds\int_0^t ds' C(s-s')$. The limiting accumulated phase is therefore Gaussian, with the Gaussian characteristic function \cite{Gardiner_2009}
\begin{equation}
\Phi_{\rm G}(\chi,t)
=
\exp\!\left[
-\frac{\chi^2}{2}
\int_0^t ds\int_0^t ds' 
C(s-s')
\right].
\label{supp-gaussian-functional}
\end{equation}

For the exponential covariance of the telegraph family, one has
\begin{equation}
\int_0^t ds\int_0^t ds' 
e^{-\gamma|s-s'|}
=
\frac{2}{\gamma^2}
\left(\gamma t-1+e^{-\gamma t}\right),
\label{supp-ou-integral}
\end{equation}
so
\begin{equation}
\Phi_{\rm OU}(\chi,t)
=
\exp\!\left[
-\frac{\chi^2\sigma^2}{\gamma^2}
\left(\gamma t-1+e^{-\gamma t}\right)
\right].
\label{supp-ou}
\end{equation}
The stationary Gaussian process with covariance $\sigma^2e^{-\gamma|\tau|}$ is the Ornstein-Uhlenbeck process
\cite{Gardiner_2009}.

Writing $u=\chi^2\sigma^2/\gamma^2$ gives the exact expansion
\begin{equation}
\Phi_{\rm OU}(\chi,t)
=
e^u
\sum_{k=0}^{\infty}
\frac{(-u)^k}{k!}
e^{-(\gamma u+k\gamma)t}.
\label{supp-ou-expansion}
\end{equation}
At fixed finite $\chi$, its exponential rates differ by integer multiples of $\gamma$, so no finite-coupling polynomial transient is generated. This is consistent with the independent-copy EP moving outward as $\sqrt M$: the Gaussian limit is taken at fixed finite $\chi$ and $t$, while the EP of the composite noise source recedes to infinite coupling.

\section{Atomtronic-ring specialization}
\label{sec:supp-ring}

We now derive the atomtronic readout used in the main text. The Hamiltonian in Eq. (\ref{ring-model}) is the standard minimal-coupling form for a synthetic gauge field acting on neutral atoms \cite{Dalibard_2011,Chen_2018}. Writing $\varphi(t)=\varphi_0+\alpha\xi(t)$, 
$L=\sum_j(-i\partial_{\theta_j})$, $E_R = 1/(2mR^2)$, expanding about the static flux gives
\begin{equation}
H(t) =
H(\varphi_0)-2E_R\alpha\xi(t)L+
NE_R\left[
2\varphi_0\alpha\xi(t)
+\alpha^2\xi(t)^2
\right],
\label{supp-ring-expand}
\end{equation}
where $H(\varphi_0) = E_R\sum_{j=1}^{N}
\left(-i\partial_{\theta_j}-\varphi_0\right)^2
+H_{\rm int}$. At fixed particle number $N$, the term proportional to $N$ is a scalar operator and therefore contributes only a realization-dependent common phase. Rotational invariance gives $[H(\varphi_0),L]=0$, so the remaining time-dependent operator commutes with the static Hamiltonian and with itself at different times. The propagator therefore factorizes as
\begin{equation}
U_\xi(t)
=
e^{-iH(\varphi_0)t}
e^{ i\Theta_\xi(t)L}
e^{-i\varphi_{\rm sc}(t)},
 \quad \quad
\Theta_\xi(t)
=
2E_R\alpha\int_0^t ds \xi(s),
\label{supp-ring-propagator}
\end{equation}
where the scalar phase $\varphi_{\rm sc}(t)$ drops out of the expectation values.

We write the density Fourier component as
$\hat n_q=\sum_j e^{iq\theta_j}$. Since
$[L,\hat n_q]=q\hat n_q$, the Baker-Campbell-Hausdorff formula gives
\begin{equation}
e^{-i\Theta L}\hat n_qe^{i\Theta L}
=
e^{-iq\Theta}\hat n_q .
\label{supp-ring-commutator}
\end{equation}
Combining Eq. (\ref{supp-ring-commutator}) with Eq. (\ref{supp-ring-propagator}) gives, for each noise
realization, the expectation value
\begin{equation}
\langle\hat n_q(t)\rangle_\xi
=
e^{-i 2E_R\alpha q
\int_0^t ds \xi(s)}
\langle\hat n_q(t)\rangle_0 .
\label{supp-ring-realization}
\end{equation}
Averaging over the noise produces Eq. (\ref{ring-fringe}) of the main text, with $\chi_q=2E_R\alpha q$. Specializing instead to the symmetric-telegraph family $\xi_M$ and setting $2\sigma|\chi_q|=\gamma\sqrt M$ gives Eq. (\ref{telegraph-hierarchy}) for the normalized density harmonic.

\end{document}